\documentclass[9pt,twocolumn,twoside]{pnas-new}

\usepackage{booktabs}
\usepackage{tabularx}

\templatetype{pnasresearcharticle} 

 \newcommand{\bit}[1]{\textit{\textbf{#1}}}
 \newcolumntype{Y}{>{\centering\arraybackslash}X}

\usepackage{xr}
\makeatletter

\newcommand*{\addFileDependency}[1]{
\typeout{(#1)}
\@addtofilelist{#1}
\IfFileExists{#1}{}{\typeout{No file #1.}}
}\makeatother

\newcommand*{\myexternaldocument}[1]{%
\externaldocument{#1}%
\addFileDependency{#1.tex}%
\addFileDependency{#1.aux}%
}

\myexternaldocument{SupplementaryMaterials}

\usepackage{hyperref}
\hypersetup{
    colorlinks=true,
    linkcolor=blue,
    filecolor=magenta,      
    urlcolor=cyan,
}

\title{Interfacial Stresses on Droplet Interface Bilayers Using Two Photon Fluorescence Lifetime Imaging Microscopy} 
\author[a, g]{Yaoqi Huang}
\author[a, b, c, g]{Vineeth Chandran Suja}
\author[a]{Menghao Yang}
\author[d]{Andrey V. Malkovskiy} 
\author[a]{Arnuv Tandon}
\author[e, f]{Adai Colom}
\author[a]{Jian Qin}
\author[a, 1]{Gerald G. Fuller}

\affil[a]{Department of Chemical Engineering, Stanford University, Stanford, CA 94305, USA}
\affil[b]{School of Engineering and Applied Sciences, Harvard University, MA - 02134, USA}
\affil[c]{Wyss Institute for Biologically Inspired Engineering, 3 Blackfan Circle, Boston, 02115}
\affil[d]{Carnegie Institute for Science, Department of Plant Biology, Stanford CA 94305, USA}
\affil[e]{Biofisika Institute (CSIC, UPV/EHU), 48940 Leioa, Spain}
\affil[f]{Department of Biochemistry and Molecular Biology, Faculty of Science and Technology, Campus Universitario, University of the Basque Country (UPV/EHU), 48940 Leioa, Spain}
\affil[g]{Equal contribution}

\leadauthor{Huang}

\significancestatement{Understanding the behavior of phospholipid membranes under mechanical stimuli is important for fundamental cell biology and for developing novel cellular therapies. Existing tools such as AFM and optical tweezers are either invasive or lack the spatiotemporal resolution required to study phospholipid mechanics. The reported framework addresses this gap by integrating two-photon lifetime imaging of an interfacially active molecular flipper with droplet interface bilayers (DIBs). The precise mechanical manipulation capabilities of DIBs together with the real time measurement of spatiotemporally resolved membrane stresses opens up numerous possibilities for studying phospholipid  membrane behavior under mechanoperturbations.}

\authorcontributions{YH, VCS, AC, JQ and GGF discussed the scientific challenges, questions and hypotheses; YH, VCS, and GGF designed the experiments; YH and AVM conducted the experiments; MY and JQ performed the MD simulations, YH, VCS, AT, MY, JQ and SM analyzed the results; YH, VCS and GGF wrote the paper.}

\authordeclaration{None of the authors have any conflict of interest.}
\correspondingauthor{\textsuperscript{1}To whom correspondence should be addressed. E-mail: ggf@stanford.edu}

\keywords{Bilayers | Molecular flippers | Interfacial Mechanics | FLIM | Two photon microscopy}

\begin{abstract}
Response of lipid bilayers to external mechanical stimuli is an active area of research with implications for fundamental and synthetic cell biology. However, there is a lack of tools for systematically imposing mechanical strains and non-invasively mapping out interfacial (membrane) stress distributions on lipid bilayers. In this article, we report a miniature platform to manipulate model cell membranes in the form of droplet interface bilayers (DIBs), and non-invasively measure spatio-temporally resolved interfacial stresses using two photon fluorescence lifetime imaging of an interfacially active molecular flipper (Flipper-TR). We established the effectiveness of the developed framework by investigating interfacial stresses accompanying three key processes associated with DIBs: thin film drainage between lipid monolayer coated droplets, bilayer formation, and bilayer separation. Interestingly, the measurements also revealed fundamental aspects of DIBs including the existence of a radially decaying interfacial stress distribution post bilayer formation, and the simultaneous build up and decay of stress respectively at the bilayer corner and center during bilayer separation. Finally, utilizing interfacial rheology measurements and MD simulations, we also reveal that the tested molecular flipper is sensitive to membrane fluidity that changes with interfacial stress - expanding the scientific understanding of how molecular flippers sense stress. 

\end{abstract}

\dates{This manuscript was compiled on \today}
\doi{\url{www.pnas.org/cgi/doi/10.1073/pnas.XXXXXXXXXX}}

\begin{document}
\maketitle
\thispagestyle{firststyle}
\ifthenelse{\boolean{shortarticle}}{\ifthenelse{\boolean{singlecolumn}}{\abscontentformatted}{\abscontent}}{}

\dropcap{T}he phospholipid membrane plays a crucial role in the structure and function of cells. The physical, chemical and biological properties of the cell membrane are actively studied using {\it in vivo} and {\it in vitro} cell models \cite{gambale1982properties, ter1993interaction,villar2011formation}. Droplet interface bilayer (DIB), a novel {\it in vitro} cell model, is a  bilayer formed between two aqueous droplets coated with lipid monolayers in a non-polar phase. DIBs are attractive due to their ability to intricately control and visualize bilayer composition and dynamics \cite{bayley2008droplet, huang2022physicochemical}. To date, DIB has been used to evaluate physical characteristics  of lipid bilayers \cite{huang2022physicochemical,dixit2012droplet,guiselin2018dynamic,huang2022influence}, electrical characteristics \cite{el2019new,punnamaraju2012triggered}, trans-membrane transport characteristics  \cite{michalak2013effect,lee2018static,gehan2020penetratin, harriss2011imaging,punnamaraju2012triggered},  and cell functionalization in a variety of conditions \cite{thutupalli2011bilayer,najem2015activation,freeman2016mechanoelectrical}.

The interfacial (membrane) stress, defined as the force per unit length acting on the lipid bilayer, is crucial to many biological processes \cite{pontes2017membrane,morris2001cell,raucher2000cell}. Cell adhesion, a fundamental property that is necessary for cell migration and multicellularity, is influenced by membrane stress \cite{maitre2013three,parsons2010cell}. During cell division, membrane tension impacts the formation of the daughter cells, with an increased tension delaying abscission - the last step of cell division  \cite{colom2018fluorescent,lafaurie2013escrt}. Membrane stresses also play a critical role in phagocytosis \cite{masters2013plasma} and can induce disruptions in the cell membrane - both of which have implications in the development of cell therapies \cite{shields2020cellular, chakrabarty2022microfluidic, stewart2016vitro}. Motivated by these processes, researchers have actively studied the effects of cell membrane stresses, including the work from this laboratory using DIBs to probe the connection between bilayer separation mechanics and membrane tension \cite{huang2021surface}. Despite the progress, it is currently not possible to obtain spatio-temporally resolved stress distributions in DIBs. Although techniques such as AFM allow the measurement of membrane stresses at a fixed location on a bilayer, none of the existing techniques are suitable for dynamically evolving bilayers \cite{benz2004correlation,garcia2010nanomechanics}.

\begin{figure*}[!h]
\centering
     \includegraphics[width=\linewidth]{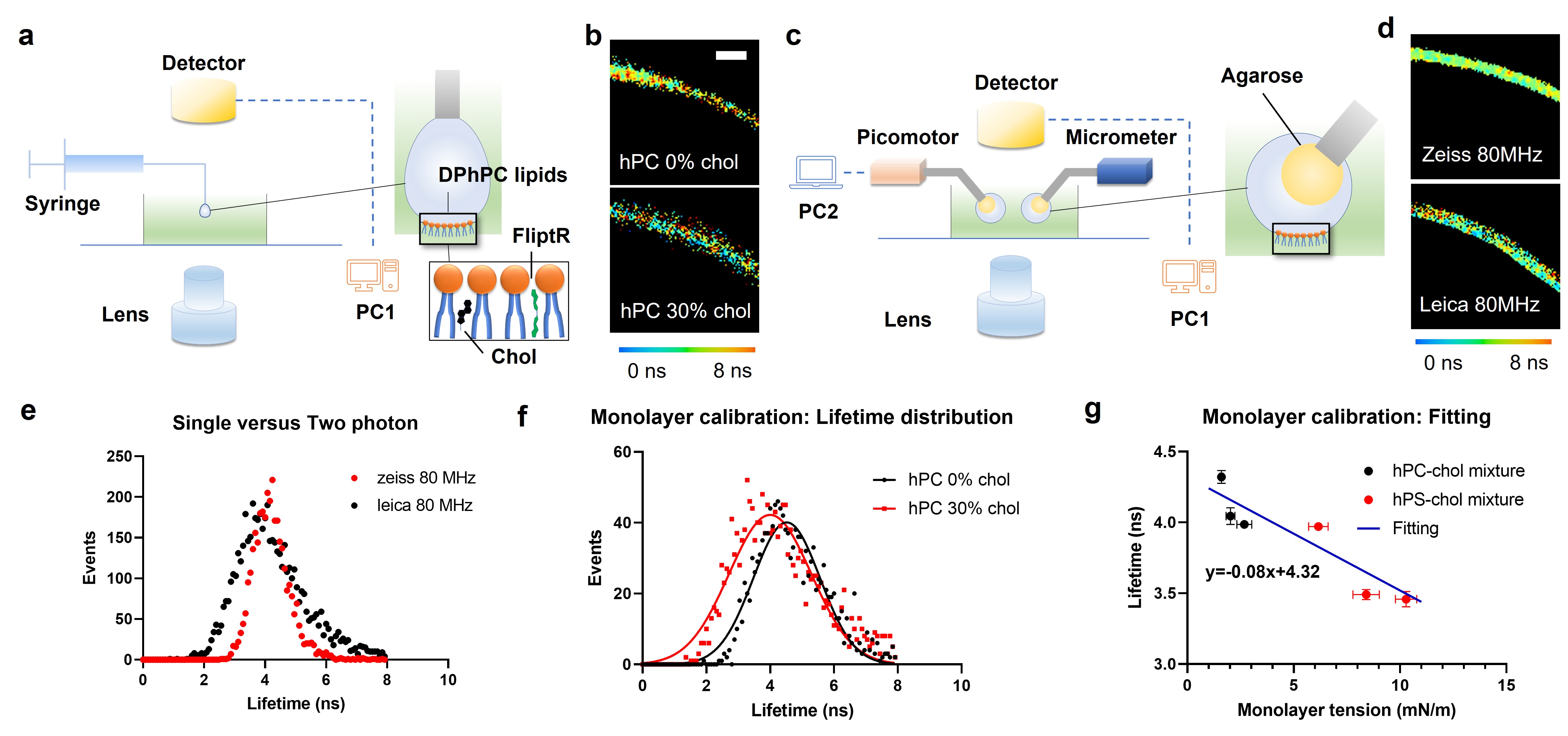} \caption{A schematic of the experimental setup and calibrations. {\bf a.} FLIM setup used for lifetime - interfacial stress calibration. A pendant droplet of KCl containing Flipper-TR is created in the hexadecane solvent with DPhPC lipids and a certain percentage of cholesterol. Lipid monolayers are formed on the surface of the droplet, and the lifetime of the Flipper-TR probe embedded inside the lipid monolayer is recorded by the FLIM detector. {\bf b.} FLIM image of a section of the pendant droplet. {\bf c.} FLIM setup for DIB experiment. Two pendant droplets pinned by agarose onto needles, and a picomotor is used to position the left droplet. {\bf d.} Monolayer FLIM images obtained by two photon microscopy and one photon microscopy. {\bf e.} Gaussian curves of the lifetimes obtained via single and two-photon microscopy shows little difference.  {\bf f.} Distribution of fluorescence lifetimes with Gaussian fits for DPhPC monolayers with different cholesterol concentrations. {\bf g.} Average lifetime variation with monolayer surface tension (N=3, where N is the number of trials). Scale bar for subfigures (b) and (d) is  0.05 mm. 
} \label{fig:setup}
\end{figure*}

Recently, a new noninvasive method has been developed where membrane stresses can be measured utilizing a fluorescent lipid tension reporter (Flipper-TR), one of the first molecular flippers to specifically measure interfacial characteristics \cite{colom2018fluorescent}. Flipper-TR consists of two dithienothiophene aromatic rings (also referred to as flippers) that can twist and planarize in response to increasing lipid packing density and membrane tension \cite{licari2020twisting}. The fluorescence lifetime of the molecule depends on the dihedral angle between the flippers, with the molecule displaying higher lifetimes as the dihedral angle increases (i.e. as the molecule becomes more planar) \cite{licari2020twisting, dal2015fluorescent, colom2018fluorescent}. This promising tool and the associated protocols, however, have yet to be optimized for dynamic stress measurements and made suitable for DIBs.

In this manuscript, we report an experimental framework and platform for incorporating Flipper-TR into DIBs and evaluating spatio-temporally resolved membrane stresses under a variety of dynamic conditions. Initially, we detail the construction and validation of a miniature inverted two-photon microscope compatible platform for creating and manipulating DIBs. We subsequently use this platform to create DIBs using two parallel droplets, and characterize dynamic stresses with two-photon fluorescence lifetime imaging (FLIM) during three key processes: approach of lipid monolayer coated droplets, bilayer formation, and bilayer separation. The key findings in this study are validated and supported by mathematical modeling, interfacial rheology measurements and molecular dynamics simulations. Finally, we conclude the manuscript by discussing interesting avenues for future research.

\section*{Results}
\subsection*{Dynamic two photon DIB FLIM experiments} 
To enable the measurement of spatio-temporally resolved phospholipid membrane stresses, we built a miniature experimental DIB platform compatible with inverted microscopes for creating and manipulating droplet interface bilayers (DIBs). As shown in Fig.\ref{fig:setup}a,c, the platform consists of a glass chamber to hold the non-polar solvent (hexadecane in our case) with lipids, and two 75 degree blunt needles (ID: 0.58 mm OD: 0.81 mm) to hold the pendant drops. The two needles are placed at the opposite sites of the chamber, with one of them connected to a picomotor, and the other one to a manually operated micrometer. Each needle has an agarose gel at its tip for anchoring the pendant drops and prevent dripping. The setup is mounted atop a two photon inverted microscope with a pulsed laser (Zeiss LSM780) for performing fluorescence lifetime imaging microscopy (FLIM). Lipid coated droplets containing the fluorescent molecular flipper Flipper-TR are generated on the two needles and brought together to create DIBs. Time correlated single photon fluorescence data is then acquired with the help of SPC150N photon counting module from Becker GmbH. Spatio-temporally resolved fluorescence lifetimes are recovered using FLIMFit \cite{warren2013rapid} by first correcting the raw data with a previously obtained instrument response function (IRF) and then fitting a double exponential function (see \bit{Methods} for more details).  

We found that two photon microscopy FLIM yields better signal-to-noise ratios compared with the single photon FLIM while imaging DIBs (\bit{see SI Fig.\ref{sfig:undetectbilayer}}). Unlike micrometer scale liposomes tested in the literature \cite{colom2018fluorescent}, DIBs are comprised of millimeter scale droplets, which require the increased imaging depth of two photon microscopes \cite{benninger2013two} for adequate visualization.  We confirmed the suitability of the developed two photon microscopy protocols for FLIM, by imaging a single pendant drop via previously established single photon microscopy protocols \cite{colom2018fluorescent} and two photon microscopy protocols. As shown in Fig.\ref{fig:setup}e, nearly identical lifetime distributions are obtained in both cases (see \bit{Methods} and \bit{SI Fig.\ref{sfig:onephoton}} for more details). 

For mapping lifetimes to membrane stresses, we measured fluorescence lifetimes of monolayers of lipid coated pendant drops with varying lipids and cholesterol concentrations (Fig.\ref{fig:setup}f). The corresponding monolayer tensions were obtained via pendant drop tensiometry \cite{suja2020single, berry2015measurement}. As shown in Fig.\ref{fig:setup}g, we find that the lifetimes decrease with increasing monolayer tensions. A linear fit yields a slope of $-0.08 \;ns\;m\;mN^{-1}$, which is comparable to those reported for liposomes \cite{colom2018fluorescent} and expected from MD simulations \cite{licari2020twisting}. In other words, fluorescence lifetime is inversely correlated to the stress acting on the interface (see \bit{Methods} for more details).

\subsection*{Dynamic FLIM with Flipper-TR is sensitive to hydrodynamic stresses}
\begin{figure}[!h]
\centering
     \includegraphics[width=\linewidth]{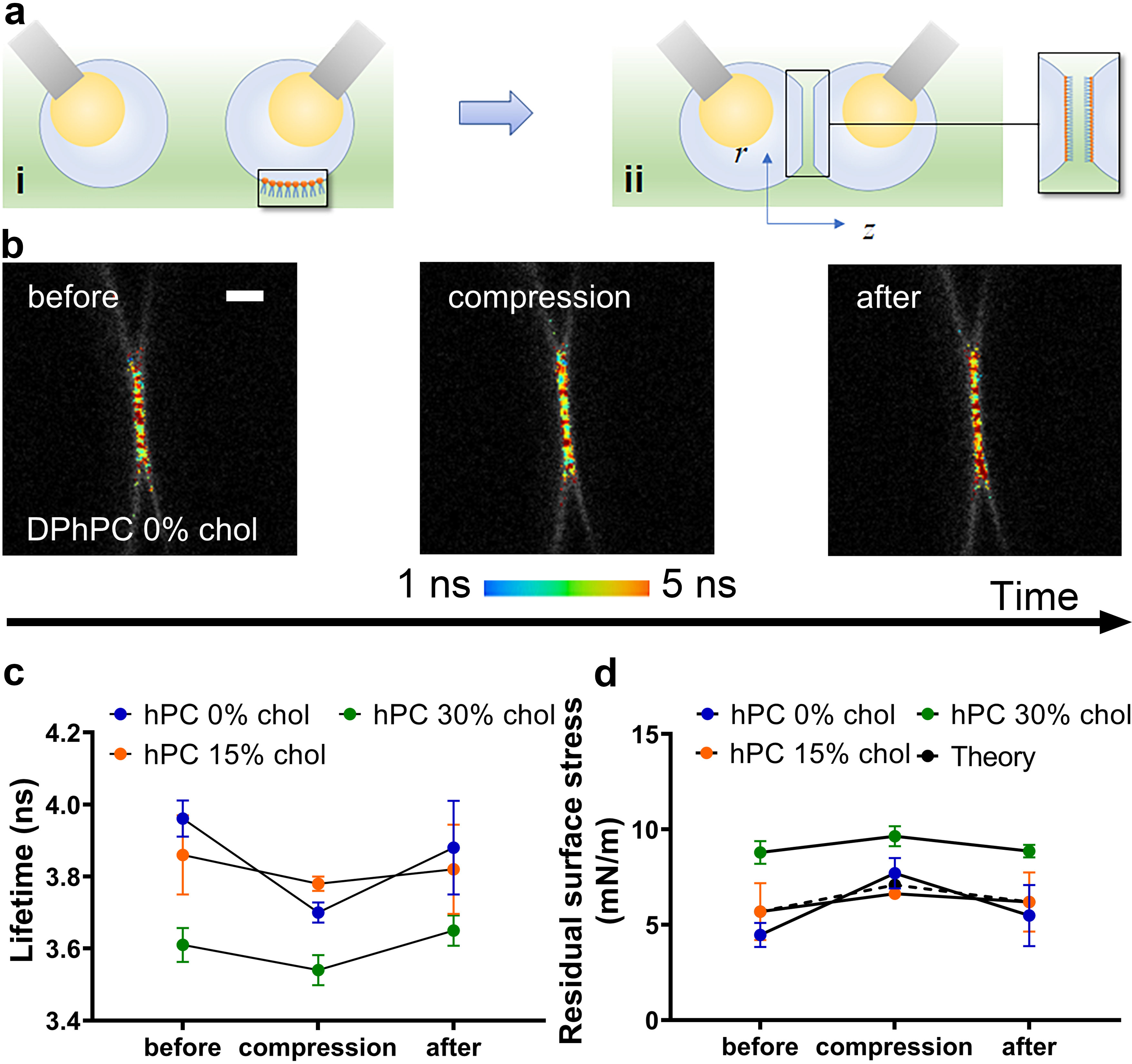} \caption{Mechanics when two monolayers are in contact. {\bf a.} A schematic of droplet profiles showing two droplets approaching against each other. {\bf b.} FLIM images showing the lifetime before, during, and after the two monolayers are in contact.  {\bf c.} Average lifetime for different lipid samples (N=3). {\bf d.} Residual surface stress along with the theoretical prediction obtained from Eq.\ref{eq:hydrodynamicStress}. Scale bar equals to 0.05 mm (N=3).
} \label{fig:monolayer in contact}
\end{figure}

The first step in creating DIBs involves pressing two lipid monolayer coated droplets (see Fig.\ref{fig:monolayer in contact}a) against one another. This process traps a thin liquid film of hexadecane between the drops, which eventually drains before bilayer formation.  

As qualitatively shown in Fig.\ref{fig:monolayer in contact}b and quantitatively in Fig. \ref{fig:monolayer in contact}c, during thin film drainage, the fluorescence lifetime of the pre-bilayer (monolayer) interface with DPhPC decreases during thin film drainage. A similar, albeit smaller decrease in lifetime, was observed in the presence of cholesterol (Fig. \ref{fig:monolayer in contact}c). We hypothesize that this dynamic change in lifetime is driven by the hydrodynamic lubrication stresses associated with the drainage of the non-polar phase.    

Assuming incompressible, Newtonian, non-polar fluid behavior and approximating the region between drops as that between two parallel disks, the hydrodynamic force per unit length ($\Pi$) acting on the monolayer can be calculated from lubrication theory as (see \bit{SI Fig.\ref{sfig:thinfilmcal}} and text for details),
\begin{equation}\label{eq:hydrodynamicStress}
    \Pi = -6 \mu \pi r^2 \frac{V_z}{h^2}.  
\end{equation}

\begin{figure*}[!h]
\centering
     \includegraphics[width=\linewidth]{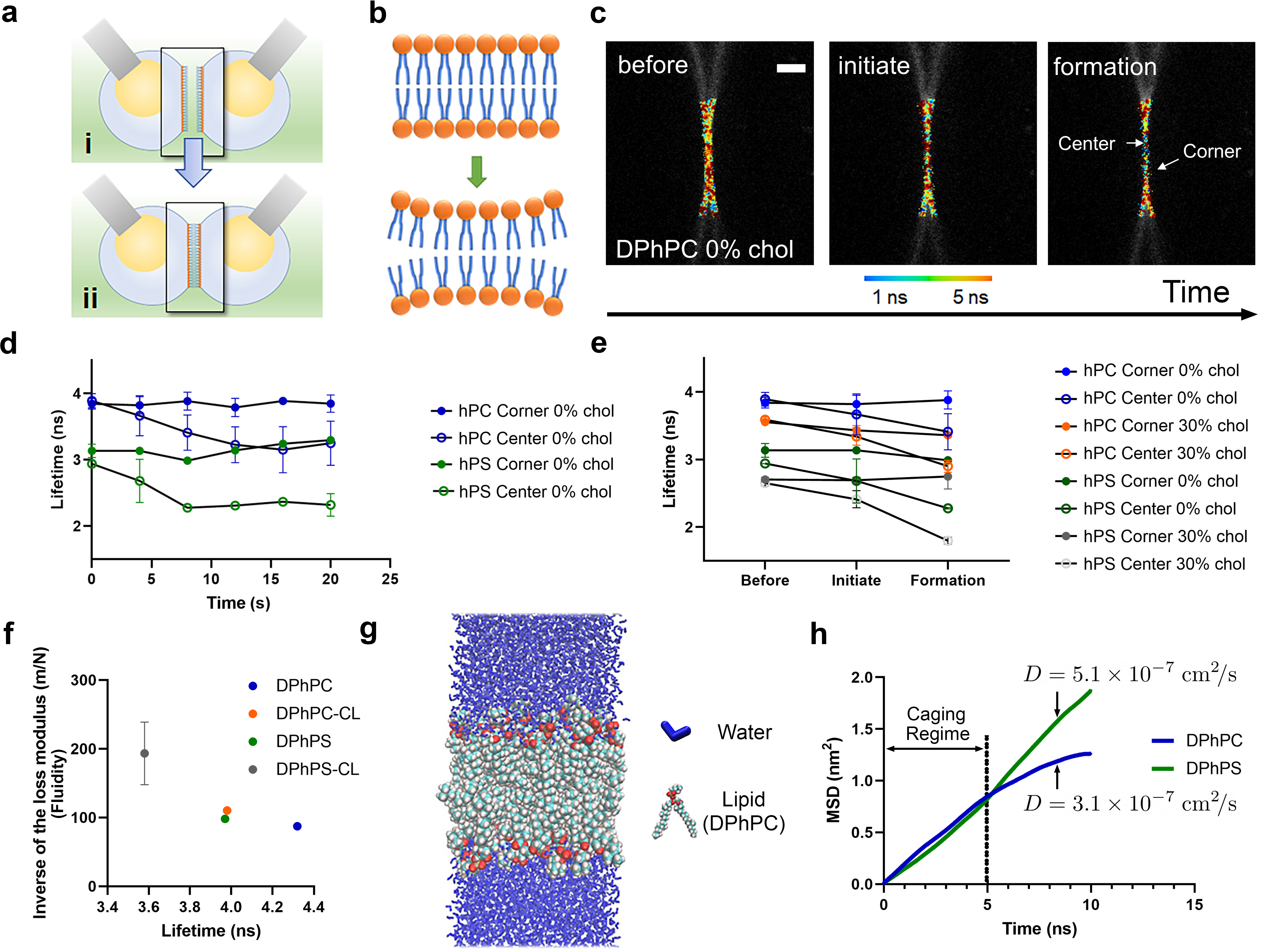} \caption{Mechanics during the bilayer formation. {\bf a.} A schematic of droplets showing the formation of the bilayer. {\bf b.} A schematic of the bilayer profile between two droplets during bilayer formation. The zipping of two monolayers changes the lipid packing. {\bf c.} FLIM images showing the lifetime of the pure DPhPC sample when two monolayers are forming a bilayer. {\bf d.} Lifetime distributions over 20s (N=3). {\bf e.} Lifetime distribution for different lipid samples (N=3). {\bf f.} Inverse of the loss modulus of PC and PS mixtures (N=3). Scale bar is equal to 0.05 mm. {\bf g.} Atomic configurations for DPhPC bilayers between water molecules. As shown in the legend, the entire water molecule is shown with a solid blue color, and the DPhPC molecule has red representing oxygen atoms, the white representing hydrogen atoms, cyan representing carbon atoms, and tan representing phosphorus atoms (not visible). {\bf h.} Mean square displacement (MSD) for DPhPC and DPhPS lipid bilayer as a function of MD simulation time. 
} \label{fig:bilayer formation}
\end{figure*}

Here $\mu$ is the non-polar bulk viscosity, $r$ is the radial co-ordinate along the monolayer, $V_z$ is the film thinning rate, and $h$ is the thickness between the monolayers. An order of magnitude analysis at a radial location of r = 0.1 mm, with typical values of relevant quantities - $\mu$ = 3 cP (for hexadecane \cite{hardy1958viscosity}), $V_z$ = 1 nm/s and h = 20 nm \cite{radoev1983critical}, yields $\Pi \sim 1.5 mN/m$. The residual stress (surface tension + hydrodynamic interfacial stress $\Pi$) on the monolayer tension during compression agrees with the estimate obtained from lubrication theory (Fig. \ref{fig:monolayer in contact}d), supporting that dynamic FLIM with Flipper-TR is sensitive to and can be used to study the effect of hydrodynamic stresses on lipid membranes. The agreement between theory and experiments also provides additional validation of the reported FLIM image acquisition, calibration and post-processing protocols.

\begin{figure*}[!h]
\centering
     \includegraphics[width=\linewidth]{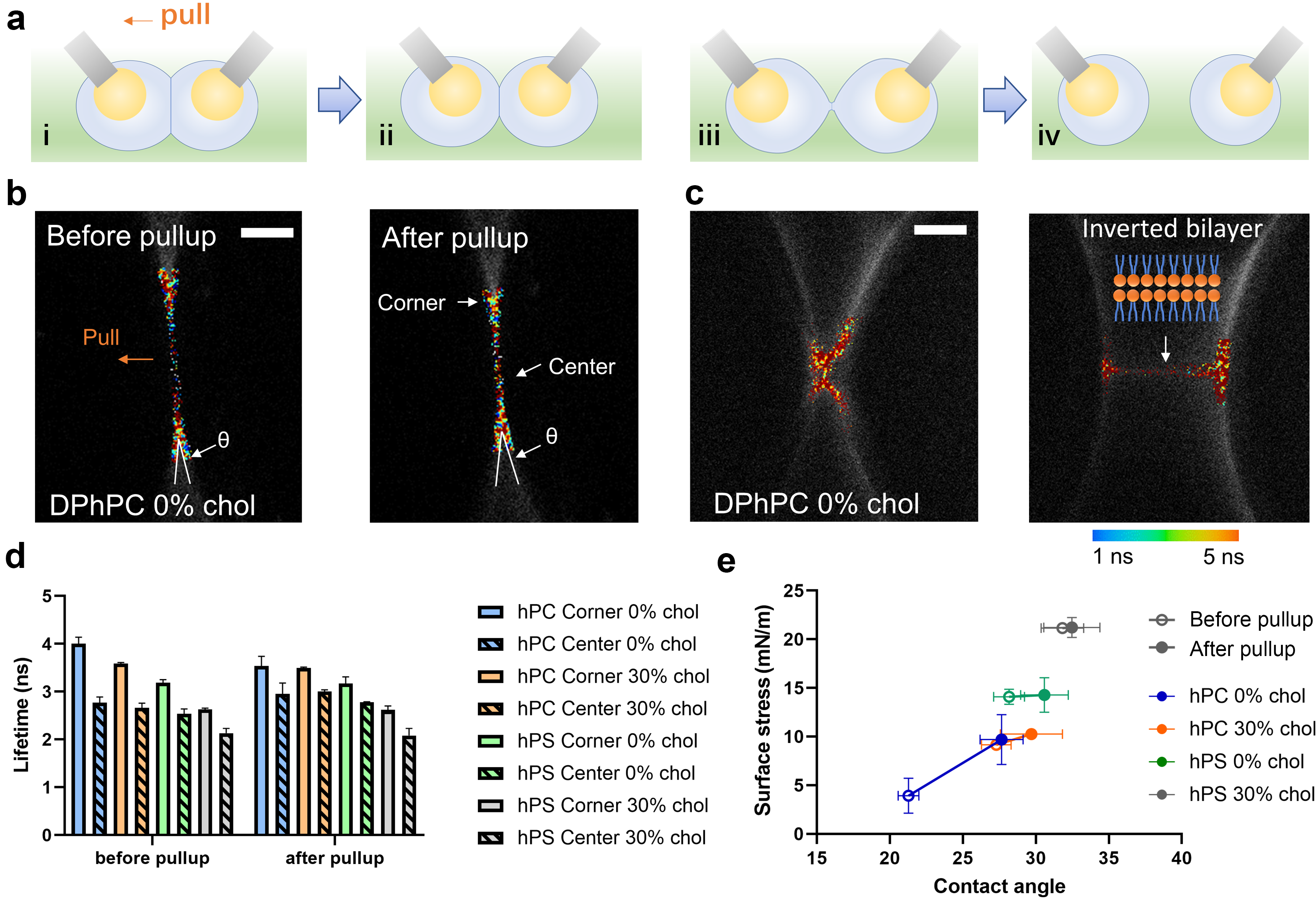} \caption{Bilayer separation mechanics. {\bf a.} A schematic of droplet profile during the peeling process (i and ii) and complete detachment (iii and iv). {\bf b.} FLIM images before and after the strain displacement. $\theta$ inside the subfigure is denoted as contact angle {\bf c.} Formation and thread extension during bilayer detachment. {\bf d.} Lifetime during the peeling process for different lipid samples (N=3).{\bf e.} Corner surface stress versus contact angle ($\theta$) before and after a step strain (pullup) for different lipid samples (N=3). Scale bar equals to 0.05 mm.
} \label{fig:bilayer peeling process}
\end{figure*}

\subsection*{Membrane stress and fluidity of droplet interface bilayers increases radially inwards}
Bilayer formation happens when hydrophobic phospholipid tails zip-up to completely exclude the oil film between the two droplets (Fig \ref{fig:bilayer formation}a,b). Supplementing the existing body of literature on the physics of bilayer formation\cite{huang2022physicochemical,vargas2014fast,thutupalli2011bilayer}, we measured the spatio-temporal fluorescence lifetime evolution during DPhPC and DPhPS bilayer formation. As shown in Fig.\ref{fig:bilayer formation}c,d, fluorescence lifetimes are inhomogeneously distributed with a lower lifetime at the bilayer center. A similar trend is observed on the addition of cholesterol (Fig.\ref{fig:bilayer formation}e) - suggesting the presence of a membrane stress field that increases inwards. Curiously, the magnitude of decrease in lifetime dramatically differs across both lipids (50\% lower in DPhPS as compared to DPhPC). This is surprising as both lipid bilayers are known to have a similar packing density (area per lipid) \cite{tomita2013does}, and Flipper-TR is expected to mechanistically respond to membrane tension via changes in lipid packing.  

To investigate this further, we first performed interfacial rheology measurements on lipid monolayers (Fig.\ref{fig:bilayer formation}f). The interfacial fluidity (the inverse of the loss modulus) scales with the lifetime. It is worth noting that addition of cholesterol increases membrane fluidity for the tested lipids, but not for DOPC - consistent with previous studies \cite{colom2018fluorescent} and likely a result of the dual role of cholesterol in regulating membrane fluidity \cite{zhang2020effect} (\bit{SI Fig.\ref{sfig: DOPClifetime}}). To confirm that the fluidity trends observed on lipid monolayers translated to lipid bilayers, we turned to MD simulations (Fig.\ref{fig:bilayer formation}g). The simulations confirmed the similar packing density of DPhPS and DPhPC lipid bilayers (\bit{SI Fig.\ref{sfig:atomarea}}). We evaluated the lipid trajectory data to compute self diffusivity - a key property that is easily obtained and correlated to membrane fluidity \cite{reddy2012effect}. As seen in Fig.\ref{fig:bilayer formation}h, DPhPS lipids have a higher diffusivity than DPhPC, and by extension DPhPS bilayers are more fluid than DPhPC bilayers - consistent with our interfacial rheology measurements. This suggests that Flipper-TR responds to changes in membrane stresses not only via changes in lipid packing, but also due to changes in membrane fluidity imparted by membrane stresses. As the Flipper-TR molecule is known to dynamically oscillate about its dihedral in the bilayer \cite{licari2020twisting}, it is not surprising that membrane fluidity is also an important physical property that dictates its lifetime. Finally, the radial variation in lifetime can be explained by circular geometry of droplet interface bilayers and radial propagation of the lipid monolayer tension acting at the edge of the bilayer (see \bit{SI Fig.\ref{sfig:radiaprop}} for details). Taken together, this suggests that droplet interfaces are more fluid as we move radially inwards. This finding has important implications in trans-membrane transport studies routinely performed with bilayers.

\subsection*{Membrane stresses increase at corners and decrease at the center during bilayer separation}
DIBs can be manipulated to apply dynamic strains for separating the bilayer. Here, following a previously established protocol\cite{huang2021surface} (see \bit{Methods}), we applied controlled step strains by separating the droplets step-wise to probe the spatio-temporal stresses during bilayer separation (Fig \ref{fig:bilayer peeling process} a, \bit{SI Fig. \ref{sfig:separationanglediameter}}). Visualizing the fluorescence lifetimes immediately before and after a single step separation (Fig.\ref{fig:bilayer peeling process} b), we find a decrease in lifetime at the corners of the bilayer. This decrease is most pronounced for DPhPC systems, while addition of cholesterol dampens this effect (Fig.\ref{fig:bilayer peeling process} d). The center lifetime is observed to mostly increase, albeit weakly, after the first step separation. Interestingly, this indicates that there is a localized increase in membrane stress at the corners and a weak relaxation of in-plane membrane stress at the center during bilayer separation. 

This evolution of the membrane stress during step separation has important implications on the evolution of bilayer contact angle (Fig.\ref{fig:bilayer peeling process}e). For identical step strains, we see that the magnitude of change in contact angle is inversely correlated to the magnitude of change in membrane stress, explaining previous observations in the literature \cite{huang2022influence}.  To physically understand this behavior, let us recall from the previous section that there is an inverse correlation between membrane stress and fluidity (Fig.\ref{fig:bilayer formation}f - h). The more fluid the membrane, the more rapid is the stress relaxation. This results in a lower change in the membrane stress post step separation, which, in turn, drives a smaller change in the contact angle due to the lower magnitude of counteracting membrane stress.    

In the terminal separation step, there is a significant increase in the center lifetime as in-plane stresses give way to normal stresses that initiate the so-called pulling mode of bilayer separation \cite{frostad2014direct,huang2021surface} (see Fig.\ref{fig:bilayer peeling process}a). Interestingly, we observe the formation of tethers (inverted bilayer tube) during the complete separation of the bilayer (Fig.\ref{fig:bilayer peeling process}c). Although tethers were only observed in DPhPC (the most rigid lipid membrane) and were not captured in other DPhPC-chol or any DPhPS samples, this is an interesting phenomenon that deserves further investigation.  

\section*{Discussion}
The behavior of phospholipid membranes under mechanical stimuli is of fundamental and practical interest, with applications in understanding biophysical processes such as cell division, cell migration and phagocytosis, and in the development of cellular therapies \cite{maitre2013three, parsons2010cell, lafaurie2013escrt, stewart2016vitro}. Existing tools are either invasive or lack the spatio-temporal resolution required to study phospholipid mechanics. To address this gap, we report a droplet interface bilayer (DIB) framework employing two-photon fluorescence lifetime imaging of an interfacially active molecular flipper (Flipper-TR) for evaluating spatio-temporally resolved membrane stresses under a variety of dynamic conditions. Two photon microscopy enhanced the imaging depth, whereas the use of interfacially active molecular flippers minimized unwanted signal from the bulk. Both of these components are vital for improving the signal-to-noise of fluorescence lifetime images of the bilayer sandwiched between millimeter scale droplets. 

Systematic experiments with DIBs established the effectiveness of the framework in resolving interfacial stresses across diverse conditions - during thin film drainage between droplets prior to bilayer formation, during bilayer formation, and in the course of bilayer separation under external stimuli. Interestingly, the experimental framework also enabled us to uncover fundamental aspects of stress distributions in DIBs. Post bilayer formation, a radially decaying membrane stress field is created within DIBs with the highest stress existing at the center. During bilayer separation under step-strain, the existing stress field becomes more uniform, with the stress gradually relaxing at the center and building up at the corners. The change in bilayer contact angle post step strain, a key metric tracked in DIB mechanoperturbation studies \cite{huang2022influence, najem2017mechanics}, is positively correlated to the change in the membrane stress. This finding explains previous reports in the literature \cite{huang2022influence} where surprisingly dissimilar changes in contact angle were observed during identical step strains in closely related DIB systems.        

Interfacially active molecular flippers such as Flipper-TR sense membrane stresses via changes in their fluorescence lifetime. The spectroscopic response and fluorescence lifetime of molecular flippers are tightly correlated to the mean dihedral angle and the twisting dynamics of the flipper about its dihedral, which in turn changes with the molecular environment. Lipid packing density changes in response to membrane stress, and has been attributed as the key molecular environment change responsible for the mechanosensitivity of Flipper-TR. Investigating the lifetimes of Flipper-TR in two phospholipid membranes (DPhPC and DPhPS) with very similar lipid packing densities, we show, supported by interfacial rheology measurements and MD simulations, that membrane fluidity can also influence the lifetime of Flipper-TR. This expands the current scientific of mechanosensation by molecular flippers. 

There are a number of avenues for extending the findings in the current work. First, existing interfacially active molecular flippers are sensitive to a number of microenvironment features such as lipid packing density and membrane fluidity, necessitating the need for time-consuming lifetime - stress calibrations on a case-by-case basis. Establishing general physical principles that can minimize the need for repeated calibrations will make this a more attractive tool. Second, a key limitation in the reported study - the absence of orthogonal membrane stress measurements - can be alleviated by incorporating tools such as optical tweezers for force measurements. Even though the calibrations based on surface tension variations are valid \cite{licari2020twisting, reddy2012effect}, the calibration range is limited by physically attainable interfacial tensions.  Incorporating tools such as optical tweezers can overcome this limitation. Finally, the ability of molecular flippers to sense membrane fluidity opens up possibilities for their use as an interfacial rheology tool. Overall, the reported results open up promising possibilities for non-invasive phospholipid stress measurements and drive advances in fundamental cell biology and in the development of novel cellular therapies.

\matmethods{
\subsection*{Materials}
DPhPC (1,2-diphytanoyl-sn-glycero-3-phosphocholine, a neutral charged lipid) and DPhPS (1,2-diphytanoyl-sn-glycero-3-phospho-L-serine, a negative charged lipid), and DOPC (1,2-dioleoyl-sn-glycero-3-phosphocholine, a neutral charged lipid) was used as model lipids for generating phospholipid bilayers reported in this manuscript. Cholesterol from ovine wool (Catalog no: 700000; Avanti Polar Lipids Inc., Alabaster, Alabama) was purchased in powder form and was then transferred and stored in chloroform solution. Prior to the start of the experiments, DPhPC (Catalog no: 850356; Avanti Polar Lipids Inc., Alabaster, Alabama), DPhPS (Catalog no: 850408; Avanti Polar Lipids Inc., Alabaster, Alabama) and cholesterol (if needed) was extracted from the suspending chloroform solution in two steps. Initially, a predetermined amount of lipid in chloroform and a certain amount of cholesterol in chloroform are mixed. Then, the chloroform in the mixed solution was evaporated off by gently blowing a stream of nitrogen for $3$ minutes. Subsequently, the residue was vacuum dried for another $30$ minutes. The chloroform free lipids and cholesterol were then dissolved in hexadecane to give a final DPhPC concentration of 10 mM with certain mole percentage of cholesterol. 

The molecular flipper probe, Flipper-TR (Catalog no: CY-SC020, Cytoskeleton Inc), was prepared as follows. The probe was initially reconstituted in 50 $\mu$l DMSO to form 1 mM solution. 1 $\mu$L of the reconstituted probe solution is extracted in a new container, and 49 $\mu$L of 1M KCl solution is added to obtain a 20 $\mu$M Flipper-TR solution in KCl. 

Agarose gel used as a core to support the pendent drop (Fig.\ref{fig:setup} c) was prepared as follows. $300$ mg of the agarose powder (Thermo Fisher Scientific, Catalog no: BP164100) was mixed with $10$ mL of distilled water at high temperature, and then cooled down to make it into the gel form \cite{holden2007functional,leptihn2013constructing}. 1M KCl solution was then used for preparing the aqueous sessile and pendant droplets. The agarose core size was ensured to be much smaller than the pendant drop size to avoid any undesired influence of the agarose core on the reported bilayer dynamics.

\subsection*{Fluorescence lifetime imaging microscopy}
Fluorescence lifetime imaging microscopy (FLIM) measurements were obtained with a Zeiss LSM780  - a two photon microscope with a pulsed laser (80MHz, 485/850 nm).  SPC150N from Becker GmbH is equipped with the microscope as the photon counting module, and FLIMfit is used for analyzing FLIM image \cite{warren2013rapid}. A GFP 525/50 nm filter is installed inside the module to collect the emission signal.

To start the FLIM experiment, non descanned two photon laser beam is applied with 485/850 nm of the wavelength. Then, the FLIM imaging was performed under two different protocols: 1) static FLIM and 2) dynamic FLIM. The static FLIM protocol is intended to capture membrane stresses on static monolayers, and it collects 20 million photon events for each image, over 20s. The dynamic FLIM protocol is intended to resolve the membrane stresses under dynamic conditions. Here we collect only 5 million photon events, taking approximately 4s. During the experiment, 10x objective lens is used, and brightfield image from T-PMT detector may be activated to confirm the formation of the bilayer (see \bit{SI Fig. \ref{sfig:brightfieldformation}}). 

For data analysis, a dual exponential model is applied to fit fluorescence decay data for the region of interest \cite{colom2018fluorescent}. Since similar tendencies were seen for the two lifetimes, the longest lifetime is reported for all experiments. The instrument response function (IRF) used for correcting the raw data was obtained as follows. 1 $\mu$L of the gold nanoparticle solution (Sigma Aldrich, 80 nm diameter, OD 1, stabilized dispersion in citrate buffer) was pipetted on a slide and was imaged for 30 s under the same FLIM parameters as the experiments reported in the manuscript. The relaxation curve of the photons were then saved and imported as an IRF calibration for correction.

\subsection*{Pendant drop experiments}

The setup in Fig. \ref{fig:setup}a is used to measure the lifetime of pendant drops coated with lipid monolayers. On the stage a glass chamber holds the oil solution and a 0.25 mL syringe with a 90 degree blunt capillary needle (ID: 0.43 mm OD: 0.63 mm, fixed by a stand) is used to create the pendant drop. 
 To conduct pendant drop experiments, a predetermined amount of KCl solution with Flipper-TR is collected by 1 mL syringe with 90 degree bend needle, and 0.55 mL of the lipid solution (previously sonicated for 1 hour to prevent aggregates) is dispensed onto the glass chamber. Then the syringe is placed inside the oil solution and fixed in place. After that, the syringe is gently pushed until a 1 $\mu$L pendant drop is observed from the microscope brightfield image. Subsequently, a sequence of FLIM images of the pendant drop were acquired for calibration. 
 
\subsection*{DIB experiments}
The setup in Fig.\ref{fig:setup}c is used to create and manipulate DIBs. As shown in the figure, the same glass chamber used for the pendant drop experiments is used to hold the oil phase solution, but with two 75 degree blunt needles (ID: 0.58 mm OD: 0.81 mm) to hold the pendant drop. The two needles are placed at the two opposite sites of the chamber, with one of them connected to a picomotor (Newport, a stepper motor with a rotary encoder controlled by PC1), and the other one connected to a manually operated micrometer. Each needle has an agarose gel at its tip for anchoring the pendant drops and prevent dripping. The glass chamber can be mounted atop the inverted microscope stage, and the droplet profiles are obtained by FLIM via PC2. 

For DIB formation and separation experiments, 0.55 mL of the oil solution is gently added into the chamber. Then outside the chamber 1 $\mu$L of 1M KCl solution with Flipper-TR dye (20 mM) is pipetted onto the agarose on each needle to form the pendant drop, and the micrometers on both left and right droplets are adjusted to make sure that the two droplets are able to immerse into the oil phase. Then, both droplets are put into the glass chamber, and the height of the right droplet is adjusted to make sure that the both droplets are placed along the same horizontal axis. Finally, the right droplet is moved towards the left droplet so that the edge distance between the two droplets are approximately 0.3 mm. The lens is adjusted as well to ensure the two droplets are focused. 

To form the bilayers 2 droplets are aged for 10 minutes in order to allow the formation of the lipid monolayers at the oil-water interfaces. After aging, the picomotor is used to slowly push the left drop against the right pendant drop for approximately 0.35 mm at 30 $\mu$m/s, and then held in place. The thin liquid film between the pendant and sessile droplets drains until the lipid monolayers are close enough to form a bilayer. 

To conduct the bilayer separation experiments, we again used the picomotor to pull the left drop away from the pendant drop in a step-wise manner at a velocity of 0.05 mm/s for one second. The step size ($d$) has a constant value of 0.05 mm, resulting in step strain of $d/R_a = 0.067$, where $R_a = 0.75$ mm is the apex drop curvature of the pendant drop. After each separation step, we allowed the bilayer to relax for 60 seconds. This process is continued until two pendant droplets separated completely. The entire process of bilayer formation and separation was captured by FLIM, and was subsequently analyzed utilizing FLIMfit and ImageJ for recovering membrane stresses. All experiments in this paper were performed at room temperature. A schematic diagram of the above-mentioned bilayer formation and separation under microscope is shown in Fig \ref{fig:setup} c.

\subsection*{MD simulations}
Both DPhPC (PC) and DPhPS (PS) lipid bilayer models were constructed by using the Membrane Builder from CHARMM-GUI \cite{jo2008charmm}. A total number of 72 DPhPC lipids was generated with a hydration of approximately 90 water molecules per lipid, and the DPhPC topology was obtained from Klauda et al. \cite{lim2011lipid}.  The DPhPS coordinates and topology were modified based on a combination of existing the PS head group and lipid chain topologies for the CHARMM36 force field. We choose the sodium cations as the counter-ions to neutralize the DPhPS lipid bilayers. Each membrane model is consisting of 72 lipids with a hydration of 90 water molecules per lipid. The initial membrane models have a dimension of 54 nm × 54 nm × 102 nm with a total of 30528 atoms for DPhPC lipid bilayers and a total of 30096 atoms for DPhPS lipid bilayers. 

After the membrane systems were constructed, we performed all-atom Molecular Dynamic (MD) simulations of DPhPC and DPhPS lipid bilayers by using NAMD software with the CHARMM36 lipid force field \cite{yu2021semi} and the modified TIP3P water model \cite{jorgensen1983comparison}. The initial membrane systems were firstly minimized to eliminate the bad atomic contacts, and then were simulated for 600 ns at 298 K. Periodic boundary conditions were performed in the X, Y, and Z directions of the membrane systems. During the MD simulations, a tetragonal unit cell for the membrane systems was maintained to keep an equal dimension between X and Y directions along the membrane plane, while Z direction changed independently under the NPT ensemble. Langevin dynamics were applied to maintain the constant temperature of 298 K for each membrane system, and the Langevin-piston algorithm was applied to maintain the constant pressure of 1.01325 bar, i.e., atmospheric pressure at sea level. As an efficient full electrostatics method, the particle-mesh Ewald (PME) method calculated the long-range electrostatic interactions, and the grid sizes were 60 × 60 × 105 for the membrane systems. A time step of 2 fs was used for MD simulations, and atomic trajectories were collected every 100 ps for the subsequent statistical analysis. When the MD simulations reached an equilibrium state, the lipid molecule trajectories were used to calculate the membrane diffusivity (see \bit{Supplementary Materials} for more details).

\subsection*{Interfacial rheology}
Interfacial shear rheology of lipids at the oil-water (KCl) interface was measured using a Discovery HR-3 rheometer (TA Instruments) with a Du Noüy ring made of platinum/iridium wires (CSC Scientific, Fairfax, VA, catalog no. 70542000) \cite{huang2022influence}. Before each experiment, the Du Noüy ring was rinsed with ethanol and water and flame treated to remove organic contaminants. Before the experiments, 10 mM lipid in oil followed by 1M KCl solution was filled within a double-wall Couette flow cell with an internal Teflon cylinder and an external glass beaker. The Du Noüy ring was carefully lowered and positioned at the oil-water interface. A time sweep was performed at a strain of 1\% (within the linear regime) and a frequency of 0.05 Hz (sufficiently low such that the effects due to instrument inertia will not be significant). The loss modulus result was recorded at 5 minutes following the creation of the oil-water interface. 

}
\showmatmethods{} 

\acknow{We thank Youngbin Lim from the Cell Science Image Facility for his help on the FLIM and Andrea Merino from Biofisika Institute for the help on the FLIMfit software. AC acknowledges funding from MCIU, PID2019-111096GA-I00; MCIU/AEI/FEDER MINECOG19/P66, RYC2018-024686-I, and Basque Government IT1625-22}
\showacknow{} 

\bibliography{References}

\end{document}



\maketitle

\SItext
\section{Two photon versus One Photon Imaging of the Bilayer}
\begin{figure}[!h]
\centering
     \includegraphics[width=\linewidth]{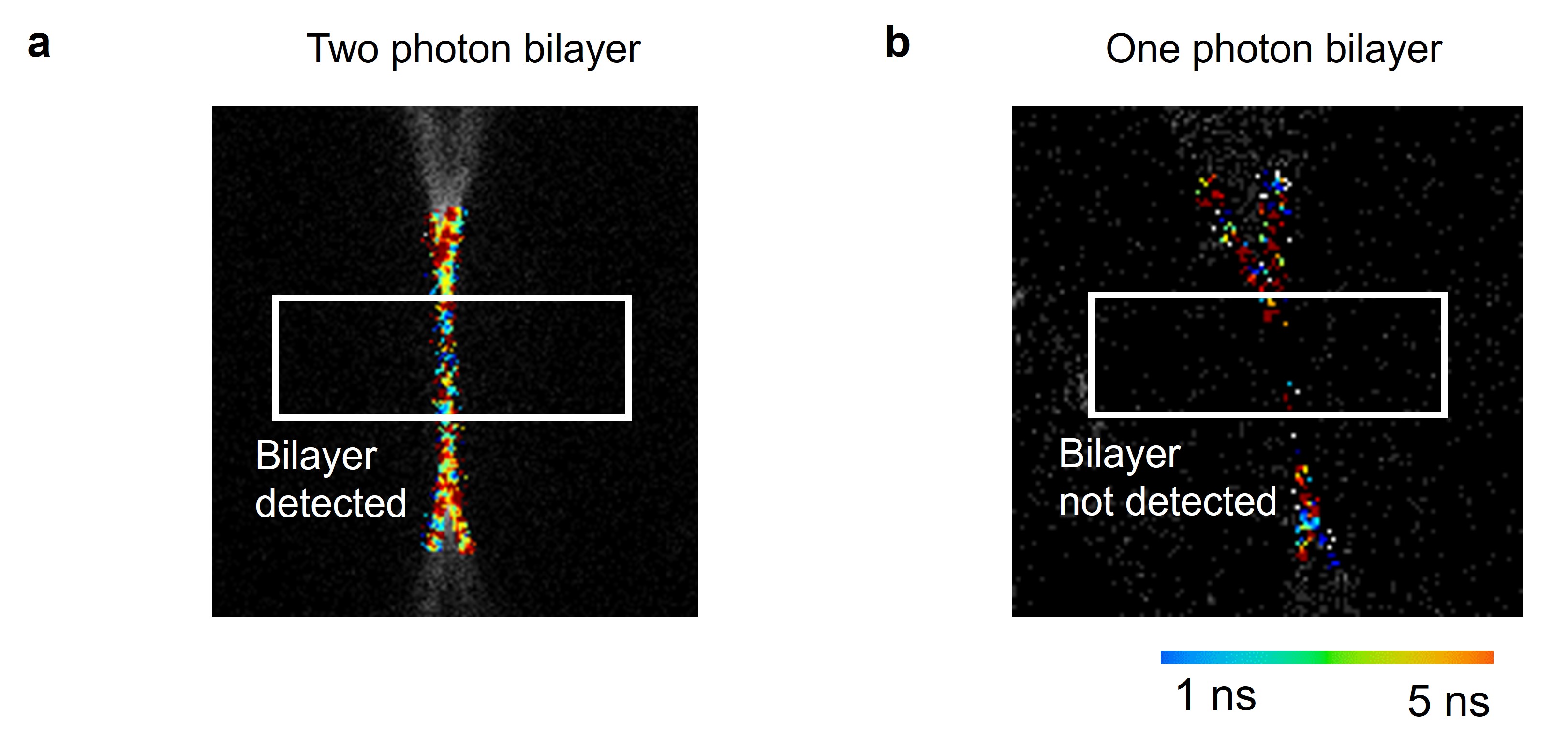} \caption{Bilayer imaged by two photon (a) and one photon (b). Fluorescence lifetime is detected by two photon imaging, but not detected by one photon microscopy.} \label{sfig:undetectbilayer}
\end{figure}

Bilayers are imaged using two photon and one photon technique under same acquisition time (Fig \ref{sfig:undetectbilayer}). We can see that the bilayer is detected under two photon but undetected using one photon microscopy, indicating that the two photon technique has advantages of imaging a droplet interface bilayer, where the thickness of the DIB makes photons hard to penetrate.

\clearpage

\section{FliptR lifetime for DPhPC lipid using one photon microscope}
\begin{figure}[!h]
\centering
     \includegraphics[width=\linewidth]{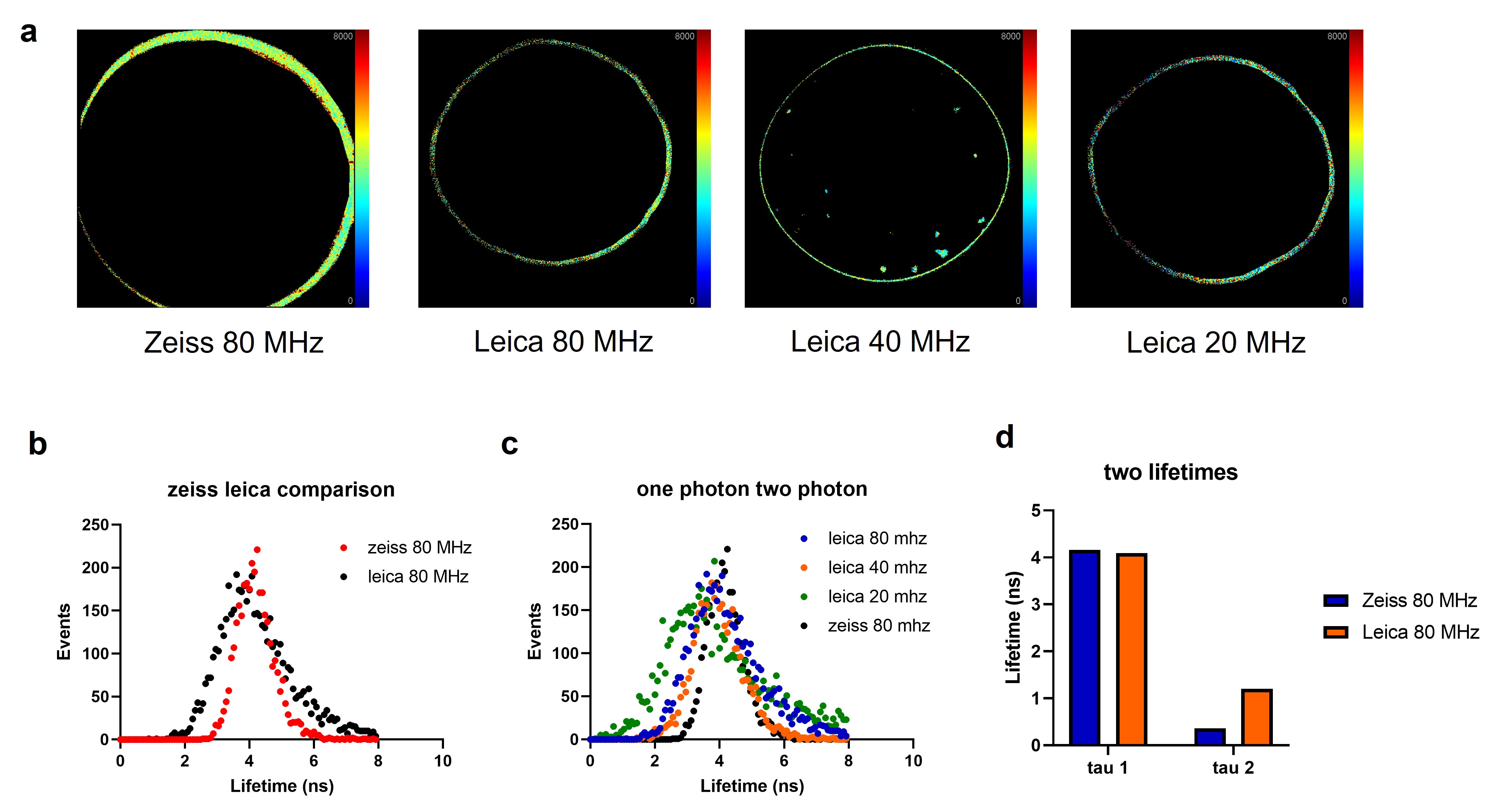} \caption{FLIM measurement using two microscopes. (a) FLIM images of monolayers on a drop pinned by agarose under Leica (one photon) and Zeiss (two photon) microscope. (b) and (c) are the Gaussian curves calculated from (a). (d) shows the long and short lifetime component for two microscopes. To keep the intensity same, 30s of the acquisition time is used from Zeiss and Leica 80 MHz, 60s is used for Leica 40 MHz and 120s is used for Leica 20 MHz.} \label{sfig:onephoton}
\end{figure}

Here, we compare the FLIM average lifetime of Zeiss microscope (two photon method) to Leica microscope (one photon method), which was used to measure the lifetime in the previous literature \cite{colom2018fluorescentSM}. The lifetime measurements using an SP8 Leica microscope were performed by a time-correlated single-photon counting module PicoQuant 300 from PicoQuant. Excitation was performed using a pulsed 488 nm white-light laser (WLL), selected by an acousto-optical beam splitter (AOBS) and a pulse-picker operating at 80, 40 and 20 MHz. The emission signal was collected through a 10x long working distance objective with an acousto-optical tunable filter (AOTF) set to the range of 490 to 580 nm. The HyD hybrid detector gain was set to 300. The FLIM measurement using Leica microscope of a single pendant drop by three laser repetition rates are shown and compared to that obtained from Zeiss. From the Fig \ref{sfig:onephoton} b and c it's clear that the lifetime results are similar, indicating the experimental data obtained from Zeiss microscope are accurate and reliable to unravel the mechanics of the DIBs. Moreover, we also calculated the shortest lifetime component tau 2 using two microscopes. We find that both of the tau 2 values at least 25\% of their corresponding values, which is also indicated in the previous literature \cite{colom2018fluorescentSM}.

\clearpage

\section{Hydrodynamic interfacial stress during thin film drainage}
\begin{figure}[!h]
\centering
     \includegraphics[width=\linewidth]{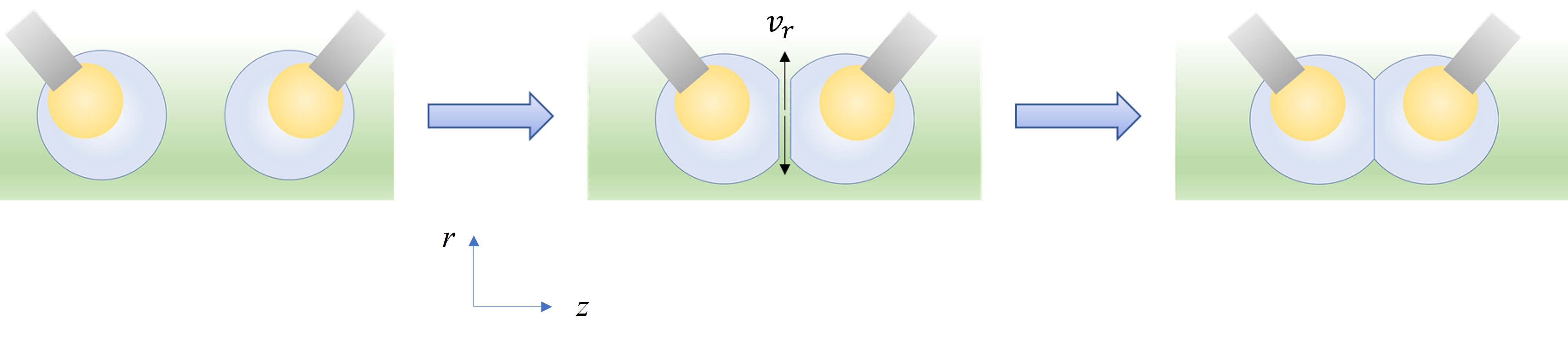} \caption{Schematic diagram for thin film theory derivation.} \label{sfig:thinfilmcal}
\end{figure}

The droplets deform and trap a thin film when they are pushed against each other (Fig. \ref{sfig:thinfilmcal}). Approximating the deformed droplet surface as a cylindrical disc, this scenario becomes identical to the squeezing flow between two plane parallel discs. The resulting radial velocity field, for no slip boundary conditions, can be written as, 

\begin{equation}
    v_r = \frac{1}{2\mu} \frac{dP}{dr} (z^2 - zh)
\end{equation}
Here $\mu$ is viscosity of the fluid in the thin film, $P$ is the pressure  and $h$ is the separation between the discs (droplets). We can solve for the pressure field within the film by first solving for $v_z$ from the continuity equation and then obtaining an ODE for pressure by applying BC for $v_z$. The solution of the resulting ODE yields, 
\begin{equation}
    P = \frac{3\mu V_z}{h^3}(R^2 - r^2) + P_0
\end{equation}
Here, $V_z$ is the rate at which the film thins in the $z$ direction, $R^2$ is the radial extent of the film and $P_0$ is the ambient pressure outside the film. Utilizing the above equations, the radial velocity gradient at the droplet surface can be expressed as follows, 

\begin{equation}
    \left. \frac{\partial v_r}{\partial z}\right|_{z = h} = -\frac{3V_zr}{h^2}
\end{equation}

\noindent The force per unit length at a radial location $r$ in follows from Newton's law of viscosity as, 
\begin{equation}
    \Pi = 2 \pi r \mu \frac{\partial v_r}{\partial z}
\end{equation}

Combining the two equations above yields:
\begin{equation}
    \Pi = -6 \mu \pi r^2 \frac{V_z}{h^2}
\end{equation}



\clearpage
\section{FliptR lifetime for DOPC lipid on single pendant droplet}
\begin{figure}[!h]
\centering
     \includegraphics[width=0.7\linewidth]{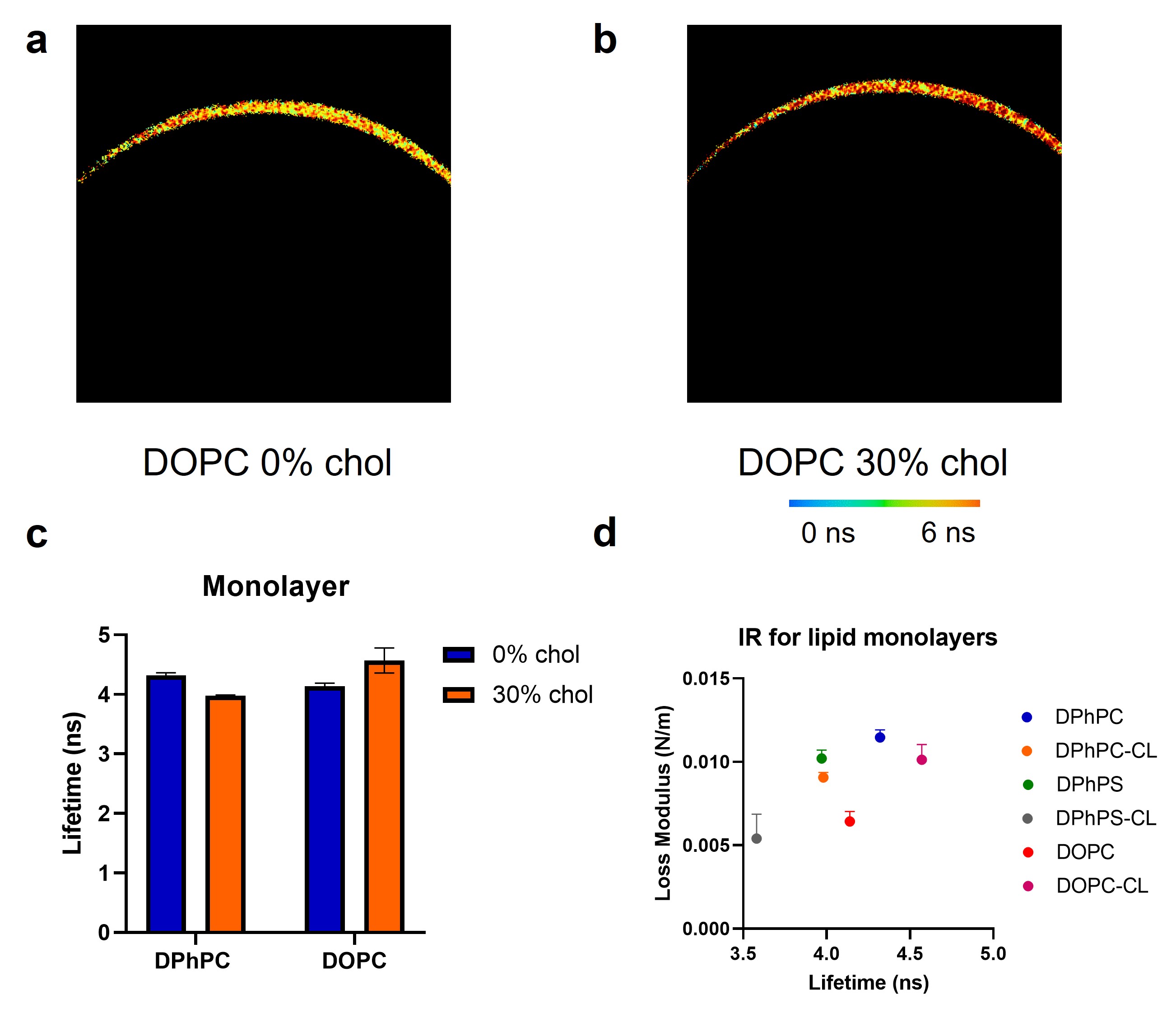} \caption{FliptR lifetime for DOPC monolayer. (a) and (b) are FLIM images. (c) shows the lifetime compared with DPhPC. (d) is the interfacial rheology (IR) data compared with other lipids.} \label{sfig: DOPClifetime}
\end{figure}

FliptR lifetime for DOPC monolayer is measured using a single pendant droplet (Fig \ref{sfig: DOPClifetime}). The lifetime is 3.98 ns for pure DOPC and 4.57 ns for DOPC with 30\% chol, indicating an increasing monolayer packing with the addition of cholesterol, which is opposite to what we have seen for DPhPC. These lifetime results and trend are comparable to the lifetime data for DOPC GUVs reported by Colom et al. \cite{colom2018fluorescentSM}, suggesting that the two photon method used for the FliptR lifetime measurement is reliable. Furthermore, we compare the loss modulus of the DOPC samples to DPhPC and DPhPS, where we find an increase of the loss modulus for DOPC with the addition of the cholesterol. The difference in behavior between DOPC and DPhPC is likely a result of the dual (bidirectional) effect of cholesterol in regulating membrane fluidity \cite{zhang2020effectSM}. 

\clearpage
\section{Methods for all-atom Molecular Dynamics Simulations}
Both DPhPC (PC) and DPhPS (PS) lipid bilayer models were constructed by using the Membrane Builder from CHARMM-GUI \cite{jo2008charmmSM}. A total number of 72 DPhPC lipids was generated with a hydration of approximately 90 waters per lipid, and the DPhPC topology was obtained from Klauda et al. \cite{lim2011lipidSM}.  The DPhPS coordinates and topology were modified based on a combination of existing the PS head group and lipid chain topologies for the CHARMM36 force field. We choose the sodium cations as the counter-ions to neutralize the DPhPS lipid bilayers. Each membrane model is consisting of 72 lipids with a hydration of 90 water molecules per lipid. The initial membrane models have a dimension of 54 nm × 54 nm × 102 nm with a total of 30528 atoms for DPhPC lipid bilayers and a total of 30096 atoms for DPhPS lipid bilayers in Fig  3 in the main text. 

After the membrane systems were constructed, we performed all-atom Molecular Dynamic (MD) simulations of DPhPC and DPhPS lipid bilayers by using NAMD software with the CHARMM36 lipid force field \cite{yu2021semiSM} and the modified TIP3P water model \cite{jorgensen1983comparisonSM}. The initial membrane systems were firstly optimized to eliminate the bad atomic contacts, and then were simulated for 600 ns at 298 K. Periodic boundary conditions were performed in the X, Y, and Z directions of the membrane systems. During the MD simulations, a tetragonal unit cell for the membrane systems was performed to keep an equal dimension between X and Y directions along the membrane plane, while Z direction changed independently under the NPT ensemble. Langevin dynamics were applied to maintain the constant temperature of 298 K for each membrane system, and the Langevin-piston algorithm was applied to maintain the constant pressure of 1.01325 bar, i.e., atmospheric pressure at sea level. As an efficient full electrostatics method, the particle-mesh Ewald (PME) method calculated the long-range electrostatic interactions, and the grid sizes were 60 × 60 × 105 for the membrane systems. A time step of 2 fs was used for MD simulations, and atomic trajectories were collected every 100 ps for the subsequent statistical analysis. 

\begin{figure}[!h]
\centering
     \includegraphics[width=\linewidth]{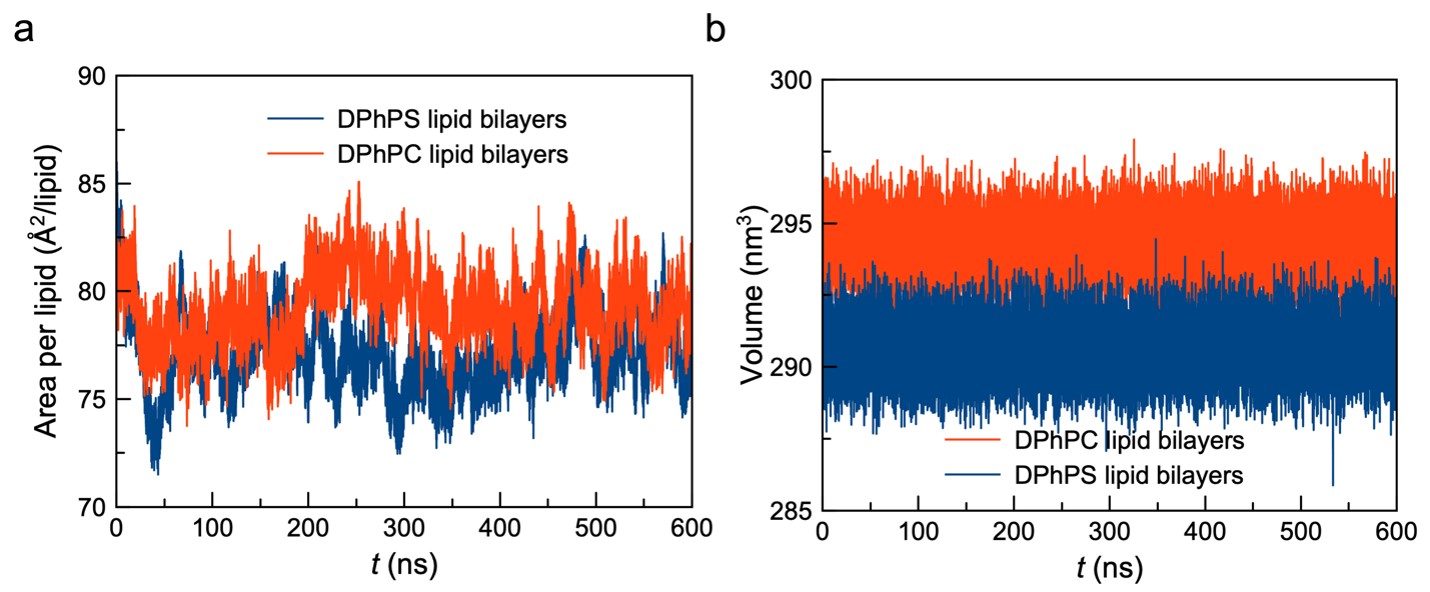} \caption{Surface area per lipid and volume for DPhPC and DPhPS lipid bilayers as a function of MD simulations time} \label{sfig:atomarea}
\end{figure}

When the MD simulations reached an equilibrate state, the atomic trajectories were used to analyze a number of important biophysical properties, such as the surface area per lipid, electrical double layer, area elastic moduli and autocorrelation functions of the surface area. As shown in Fig \ref{sfig:atomarea}, the final 350 ns of MD simulations for DPhPC lipids were used for statistical analysis, while the final 550 ns of MD simulations for DPhPS lipids were used to analyze the membrane properties.  The average surface area per lipid is a well-known property for the membrane system. The DPhPC lipids exhibited a higher average surface area of 79.36 Å$^2$ per lipid than the DPhPS lipids, with an average surface area of 77.06 Å$^2$ per lipid. Besides, the surface area per lipid of DPhPC was consistent with the previous experiments and simulations. For example, Huang et al. \cite{wu1995xSM} performed X-ray lamellar diffraction of lipid bilayer membranes and measured the surface area of DPhPC lipid bilayers as 76 Å$^2$ per lipid at 298 K. Besides, Nagle et al. \cite{tristram2010structureSM} determined the surface area of DPhPC lipid bilayers as 80.5 $\pm$1.5 Å$^2$ per lipid by using X-ray and neutron scattering data.


The diffusivity is calculated from the trajectory $r_i(t)$ of forked carbon on each lipid molecule on the bilayer \cite{he2018statisticalSM}. The displacement $\Delta \textbf{r}_i$ of a lipid molecule from time $t_1$ to $t_2$ is calculated as:

\begin{equation}
    \Delta \textbf{r}_i (t_2 - t_1) = r_i (t_2)- r_i(t_1)
\end{equation}
Then average mean squared displacement (MSD) as a function of time interval $\Delta t$ is calculated as:


\begin{equation}
   \left \langle MSD(\Delta t) \right \rangle = \frac{1}{N} \sum^N_{i=1} |\textbf{r}_i(\Delta t)-\textbf{r}_i(0)|^2
\end{equation}

Here $N$ is the total number of the forked carbons (equal to the number of lipids).  Finally, based on the Einstein relation, the diffusivity can be estimated as:
\begin{equation}
  D = \frac{ \left \langle MSD(\Delta t) \right \rangle }{2d \Delta t}
\end{equation}
where $d=2$ is the dimension of the system.

\clearpage

\section{Radial propagation of membrane tension}
\begin{figure}[!h]
\centering
     \includegraphics[width=0.4\linewidth]{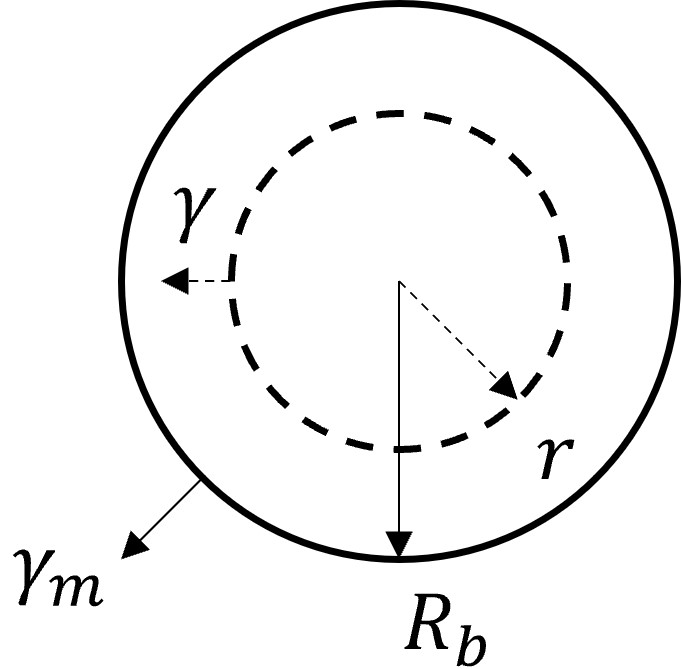} \caption{Radial propagation of the membrane tension along the bilayer radius.} \label{sfig:radiaprop}
\end{figure}

Here, we show that the stress inside the circular geometry of the interface decreases with the propagation of the bilayer radius (Fig. \ref{sfig:radiaprop}). The force acted on the bilayer interface is given by:
\begin{equation}
    F = 2 \pi r \gamma 
\end{equation}
where $F$ is the force, $r$ is the variable presenting radius and $\gamma$ is the membrane stress at a given $r$. At the edge of the bilayer, we then have:
\begin{equation}
    F = 2 \pi R_b \gamma_m
\end{equation}
where $R_b$ is the bilayer radius and $\gamma_m$ is the monolayer surface tension. Combining these equations, we have:
\begin{equation}
    2 \pi r \gamma =2 \pi R_b \gamma_m
\end{equation}
Then we obtain
\begin{equation}
    \gamma = \gamma_m \frac{R_b}{r}
\end{equation}
where the membrane stress $\gamma$ decreases with increasing $r$. 

\clearpage

\section{FLIM studies on DIB: bilayer separation}


\begin{figure}[!h]
\centering
     \includegraphics[width=\linewidth]{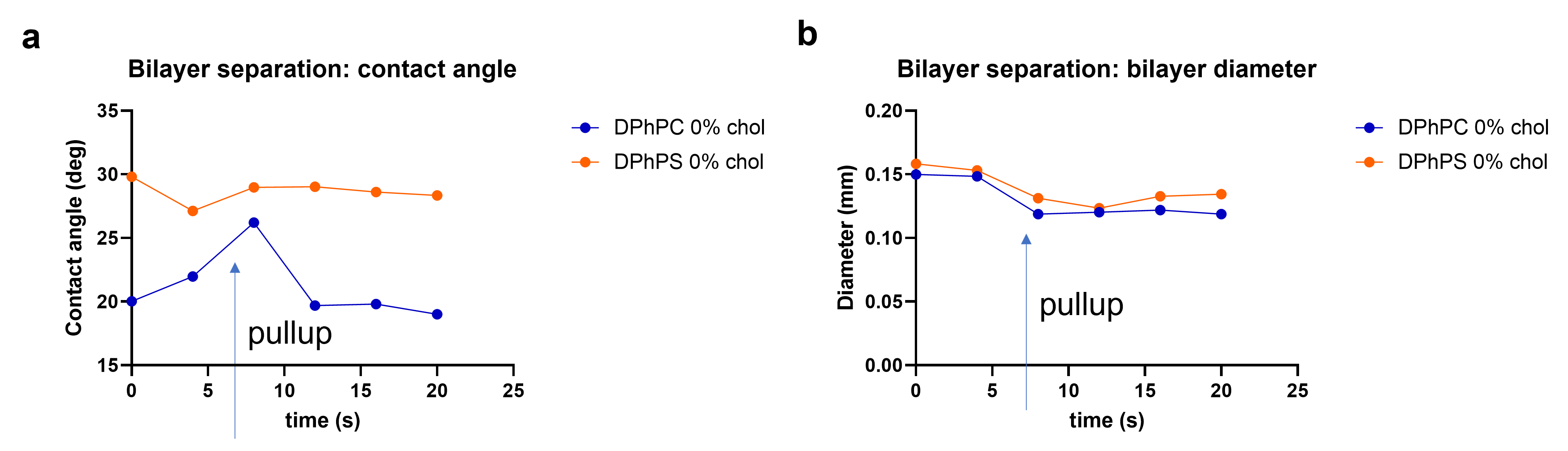} \caption{Angle and diameter change in 20s.} \label{sfig:separationanglediameter}
\end{figure}

\clearpage
\section{Brightfield image}
\begin{figure}[!h]
\centering
     \includegraphics[width=\linewidth]{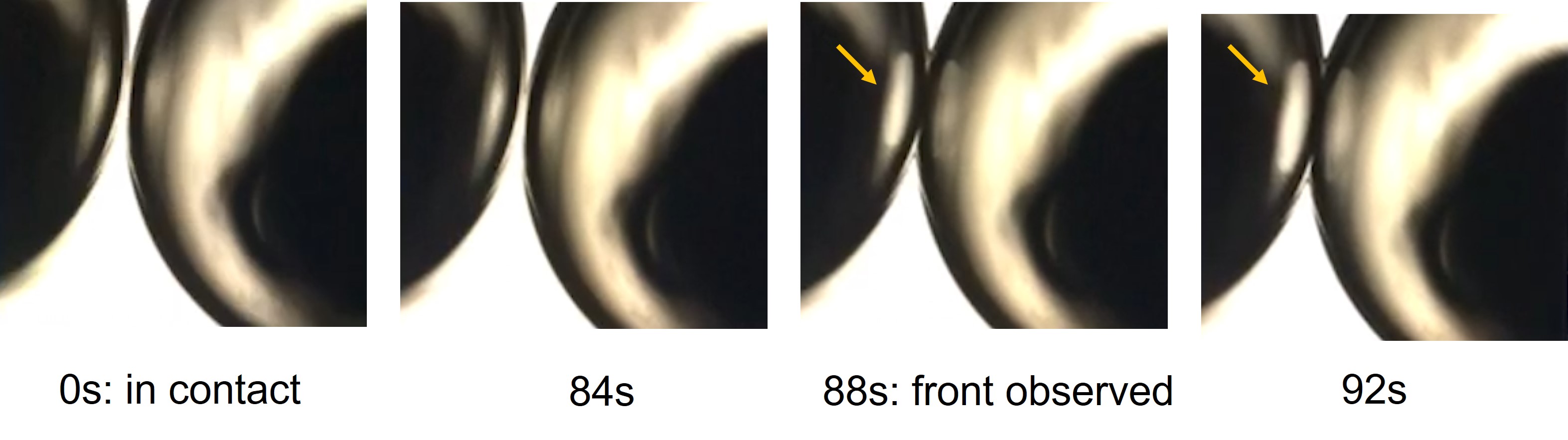} \caption{Bilayer formation confirmed by brightfield image. At 88s, a bump (shown as yellow arrow) is observed, indicating a finish of the front propagation.} \label{sfig:brightfieldformation}
\end{figure}

Bilayer formation can be confirmed via brightfield image (Fig \ref{sfig:brightfieldformation}). After the front propagation, a bump can be seen at the side of the bilayer interface.  

\clearpage

\FloatBarrier


\bibliography{SMReferences}